\RequirePackage[mathlines]{lineno}
\documentclass[prd,twocolumn,showpacs,amsmath,amssymb]{revtex4}
\usepackage{overpic,graphicx}
\usepackage{dcolumn}
\usepackage{bm}
\usepackage{rotating}
\usepackage{subfigure}
\usepackage{color}
\usepackage{multirow}
\usepackage{epstopdf}
\usepackage{mathrsfs}
\usepackage{comment}
\usepackage{booktabs}
\usepackage{hyperref}

\newcommand{\gev}{\rm \,GeV }

\newcommand{\ipb}{$\ensuremath{\mathrm{pb}^{-1}}$}
\newcommand{\ee}{e^+e^-}
\newcommand{\LCpair}{\Lambda^{+}_{c}\bar\Lambda^{-}_{c}}
\newcommand{\Lcsg}{\Lambda^{+}_{c} \to \Sigma^{+} \gamma}

\newcommand{\Bpkpi}{\bar{p}K^{+}\pi^{-}}
\newcommand{\Bpks}{\bar{p}K_{S}}

\newcommand{\LamCB}{\bar{\Lambda}_{c}^{-}}

\newcommand*{\pkpi}{\ensuremath{pK^-\pi^+}}
\newcommand*{\pks}{\ensuremath{pK_{\mathrm{S}}^{0}}}

\newcommand*{\dE}{\ensuremath{\Delta\mathrm{E}}}

\makeatletter

\newcommand{\Rmnum}[1]{\expandafter\@slowromancap\romannumeral #1@}
\makeatother

\newcommand{\mbc}{M_{\rm BC}}

\renewcommand\tablename{\rm Table}

\let\oldequation\equation
\let\oldendequation\endequation

\renewenvironment{equation}
  {\linenomathNonumbers\oldequation}
  {\oldendequation\endlinenomath}

\begin{document}

\title{\bf \boldmath
Search for the weak radiative decay $\Lcsg$ at BESIII
}

\author{
M.~Ablikim$^{1}$, M.~N.~Achasov$^{11,b}$, P.~Adlarson$^{70}$, M.~Albrecht$^{4}$, R.~Aliberti$^{31}$, A.~Amoroso$^{69A,69C}$, M.~R.~An$^{35}$, Q.~An$^{66,53}$, X.~H.~Bai$^{61}$, Y.~Bai$^{52}$
% ...
\\
\vspace{0.2cm}
(BESIII Collaboration)\\
\vspace{0.2cm} {\it
$^{1}$ Institute of High Energy Physics, Beijing 100049, People's Republic of China\\
% ...
$^{4}$ Bochum Ruhr-University, D-44780 Bochum, Germany\\
% ...
$^{11}$ G.I. Budker Institute of Nuclear Physics SB RAS (BINP), Novosibirsk 630090, Russia\\
% ...
$^{31}$ Johannes Gutenberg University of Mainz, Johann-Joachim-Becher-Weg 45, D-55099 Mainz, Germany\\
% ...
$^{35}$ Liaoning Normal University, Dalian 116029, People's Republic of China\\
% ...
$^{52}$ Southeast University, Nanjing 211100, People's Republic of China\\
$^{53}$ State Key Laboratory of Particle Detection and Electronics, Beijing 100049, Hefei 230026, People's Republic of China\\
% ...
$^{61}$ University of Jinan, Jinan 250022, People's Republic of China\\
% ...
$^{66}$ University of Science and Technology of China, Hefei 230026, People's Republic of China\\
% ...
$^{69}$ University of Turin and INFN, (A)University of Turin, I-10125, Turin, Italy; (B)University of Eastern Piedmont, I-15121, Alessandria, Italy; (C)INFN, I-10125, Turin, Italy\\
$^{70}$ Uppsala University, Box 516, SE-75120 Uppsala, Sweden\\
% ...
$^{b}$ Also at the Novosibirsk State University, Novosibirsk, 630090, Russia\\
% ...
}
%
}

\begin{abstract}

The Cabibbo-allowed weak radiative decay $\Lcsg$ has been searched for in a sample of $\LCpair$ pairs produced in $\ee$ annihilations, corresponding to an integrated luminosity of $4.5~\mathrm{fb}^{-1}$ collected with the BESIII detector at center-of-mass energies between 4.60 and 4.70$\gev$.
No excess of signal above background is observed, and we set an upper limit on the branching fraction of this decay  to be ${\mathcal B}(\Lcsg)<4.4\times10^{-4}$ at a confidence level of 90\%, which is in agreement with Standard Model expectations.
\end{abstract}


\maketitle

\oddsidemargin  -0.2cm
\evensidemargin -0.2cm

\begin{linenumbers}

\section{Introduction}
Charmed baryons provide an excellent laboratory for studying the dynamics of light quarks in the environment of a heavy quark~\cite{01, 01b}.
However, to date, there is no satisfactory phenomenological approach for describing the complicated physics of charmed-baryon decays. Improved experimental results are essential for us to understand better the underlying physics and constrain the relevant models.
In recent years, great progress
has been made in the experimental study of the  $\Lambda^+_c$ baryon at the BESIII, Belle and LHCb experiments, including  precise measurements of the $e^+e^-\to \Lambda^+_c\bar \Lambda^-_c$ production cross sections~\cite{cro}, the branching fractions (BFs) of Cabibbo-favored and suppressed hadronic decays~\cite{p1, p2a1}, the BFs of semi-leptonic decays~\cite{03a2, p2} and of hadronic weak decays~\cite{03}, as well as  searches for very rare processes~\cite{02}.
Thanks to a  large $\Lambda^+_c\bar \Lambda^-_c$ data sample, corresponding to an integrated luminosity of $4.5~\mathrm{fb}^{-1}$, BF studies at BESIII have the potential to reach a sensitivity  of  $10^{-4}$, which is sufficient to search for unmeasured  $\Lambda^+_c$  decays with low BFs and perform precise measurements of the BFs of known $\Lambda_{c}^{+}$ decays. For example, with this sample  BESIII has reported the first observation of the Cabibbo-suppressed decay $\Lambda_{c}^+ \to n \pi^+$ with a BF of $(6.6\pm1.3)\times 10^{-4}$~\cite{p3}. Throughout this paper, the charge conjugated decay channels are implied.

 Radiative decays of charmed hadrons play an important role in understanding their dynamics. Weak radiative decays usually receive contributions from both the weak and electromagnetic interactions. In addition, long-distance effects could be comparable to, or
dominant over, the short-distance ones.
In 2008, the BFs of the radiative charm decays $D^{0} \to \bar {K}^{*0} \gamma$ and $D^{0} \to \phi \gamma$ were measured by BABAR to be $(3.22 \pm 0.20 \pm 0.27)\times10^{-4}$ and $(2.73 \pm 0.30 \pm 0.26) \times10^{-5}$, respectively~\cite{p6a}, which are consistent with Standard Model predictions~\cite{08}. In 2016, Belle reported the observation of $D^{0} \to \rho^{0} \gamma$, $D^{0} \to \bar{K}^{*0} \gamma$ and $D^{0} \to \phi \gamma$ with BFs of $(1.77 \pm 0.30 \pm 0.07)\times 10^{-4}$, $(4.66 \pm 0.21 \pm 0.21)\times 10^{-4}$~\cite{p6} and $(2.76 \pm 0.19 \pm 0.10)\times 10^{-5}$, respectively. These values are  considerably larger than theoretical expectations~\cite{16}.

\begin{figure}[!htp]
\centering
\includegraphics[width=0.45\textwidth]{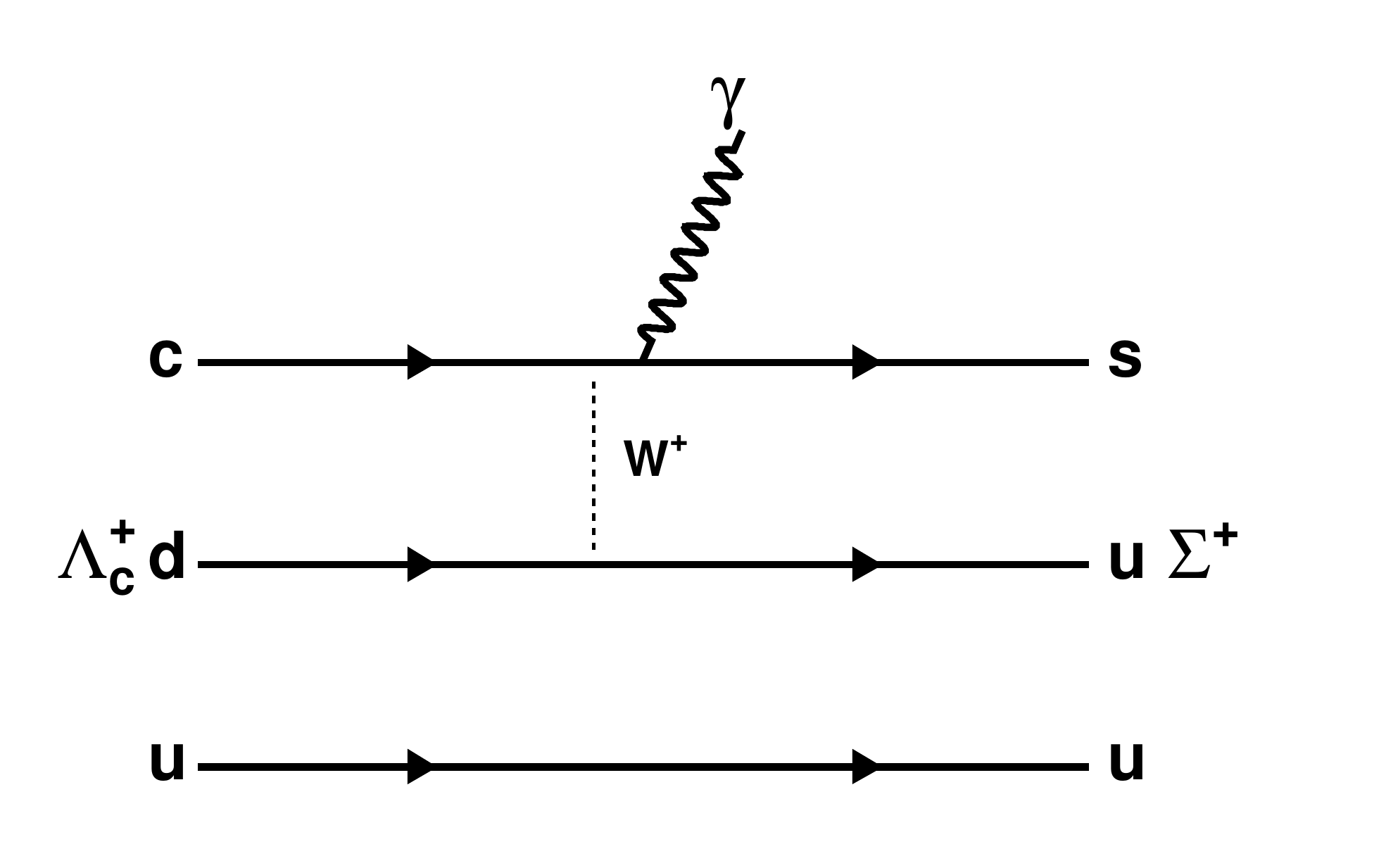}
\caption{The lowest-order Feynman diagram of the $\Lcsg$ decay.}\label{fig:fey}
\end{figure}

To date, however, the radiative decays of charmed baryon $\Lambda^+_c$ have rarely been probed in experiment.
The bag model, the constituent quark model with QCD corrections and light-cone sum rules predict the BF of the Cabibbo-allowed weak radiative decay $\Lcsg$ to be around $10^{-4}-10^{-5}$~\cite{08, qcd, light}.
The decay  proceeds predominantly through a $W$-exchange diagram
accompanied by photon emission from the external quark, as shown in Fig.~\ref{fig:fey}.
Measurement of the BFs of this kind of decay is important for distinguishing among various theoretical calculations~\cite{01, 08, qcd, light}, thereby improving the understanding of the weak radiative decay of charmed baryons.

 In June 2022, Belle reported the upper limits at 90\% credibility level on the absolute branching fraction ${\mathcal B}(\Lcsg)<2.6\times10^{-4}$~\cite{bel}. This result is consistent with the theoretical predictions~\cite{08, qcd, light}.
This paper reports the first search for $\Lambda^{+}_{c} \to \Sigma^{+} \gamma$ in a model-independent approach,
based on data sample corresponding to an integrated luminosity of 4.5~fb$^{-1}$ accumulated with the BESIII detector at the center-of-mass energies $\sqrt s= 4.60 - 4.70$~GeV~\cite{lum1}.
The center-of-mass energy and the integrated luminosity
for each energy point are listed in \tablename~\ref{tab:data_sets}.

\begin{table}[!htbp]
  \begin{center}
    \caption{The center-of-mass energy and the integrated luminosity ($\mathcal{L}_{\rm int}$)
         for each energy point.
             The first and the second uncertainties are statistical and systematic, respectively.}
    \begin{tabular}{ c  c}
      \hline
      \hline
         $\sqrt s$ (MeV) &  $\mathcal L_{\rm int}$ (\ipb)  \\
      \hline
              4599.53 $\pm$ 0.07 $\pm$ 0.74   &  586.90  $\pm$ 0.10 $\pm$ 3.90 \\
              4611.84 $\pm$ 0.12 $\pm$ 0.28   &  103.45  $\pm$ 0.05 $\pm$ 0.64 \\
              4628.00 $\pm$ 0.06 $\pm$ 0.31   &  519.93  $\pm$ 0.11 $\pm$ 3.22 \\
              4640.67 $\pm$ 0.06 $\pm$ 0.36   &  548.15  $\pm$ 0.12 $\pm$ 3.40 \\
              4661.22 $\pm$ 0.06 $\pm$ 0.29   &  527.55  $\pm$ 0.12 $\pm$ 3.27 \\
              4681.84 $\pm$ 0.08 $\pm$ 0.29   &  1664.34 $\pm$ 0.21 $\pm$ 10.32 \\
              4698.57 $\pm$ 0.10 $\pm$ 0.32   &  534.40  $\pm$ 0.12 $\pm$ 3.31 \\
      \hline\hline
    \end{tabular}
      \label{tab:data_sets}
  \end{center}
\end{table}

\section{BESIII detector and Monte Carlo}
The BESIII detector~\cite{BESCol} records symmetric $e^+e^-$ collisions
provided by the BEPCII~\cite{BEPCII} storage ring,
which operates in the center-of-mass energy ($\sqrt{s}$) range from 2.0 to
4.95~GeV,with a peak luminosity of $1 \times
10^{33}\;\text{cm}^{-2}\text{s}^{-1}$ achieved at $\sqrt{s} =
3.77\;\text{GeV}$.
BESIII has collected large data samples at these energy regions~\cite{detector1}.
The cylindrical core of the BESIII detector covers 93\% of the full  solid angle and
consists of a helium-based multilayer drift chamber~(MDC), a plastic scintillator time-of-flight
system~(TOF), and a CsI(Tl) electromagnetic calorimeter~(EMC),
which are all enclosed in a superconducting solenoidal magnet providing a
{\spaceskip=0.2em\relax 1.0 T} magnetic field~\cite{detectorY}. The solenoid is supported by an
octagonal flux-return yoke with resistive plate counter based muon
identification modules interleaved with steel.
The charged-particle momentum resolution at $1~{\rm GeV}/c$ is $0.5\%$,
and resolution of the ionization energy loss in the MDC ($\mathrm{d}E/\mathrm{d}x$)
is $6\%$ for electrons
from Bhabha scattering. The EMC measures photon energies with a
resolution of $2.5\%$ ($5\%$) at $1$~GeV in the barrel (end-cap) region.
The time resolution in the TOF barrel region is
68 ps, while that in the end-cap region is 110 ps.
The end-cap TOF system was upgraded in 2015 using multi-gap
resistive plate chamber technology, providing a time
resolution of 60 ps~\cite{detector2}.

Simulated samples produced with the {\sc geant4}-based~\cite{geant4} Monte Carlo (MC) package, which
includes the geometric description~\cite{geod1,geod2} of the BESIII detector and the
detector response, are used to determine the detection efficiency
and to estimate the backgrounds. The simulation includes the beam-energy spread and initial-state radiation (ISR) in the $e^+e^-$
annihilations modeled with the generator {\sc kkmc}~\cite{ref:kkmc}. The inclusive MC sample, which consists of $\Lambda_c^+\bar{\Lambda}_c^-$ events, $D_{(s)}$ production, ISR return to lower-mass $\psi$ states, and continuum processes ($e^{+}e^{-}\rightarrow u\bar{u},  d\bar{d}$ and  $s\bar{s}$) is generated to estimate the potential background, in which all the known decay modes of charmed hadrons and charmonia are modeled
with {\sc evtgen}~\cite{g1, g2} using BFs taken from the
Particle Data Group (PDG)~\cite{g3},
and the remaining unknown decays are modeled with {\sc lundcharm}~\cite{g4, g4a}.
Final-state radiation~(FSR) from charged final state particles is incorporated using
{\sc photos}~\cite{g5}. The $e^+e^- \to \Lambda_c \bar{\Lambda}_c$ line-shape implements the description from~\cite{cro}. The signal MC events of $\Lcsg$ are modeled with a phase-space generator.
\end{linenumbers}

\section{Methodology}
 At $\sqrt s=4.60 - 4.70$~GeV, $\LCpair$ pairs are produced in $\ee$ annihilations without additional hadrons. The $\LamCB$ baryons are fully reconstructed by their hadronic decays to $ \bar {p}K^+\pi^-$, $\bar {p}K^{0}_{S}$, $\bar {p} K^+ \pi^- \pi^0$, $\bar {p}K^{0}_{S}\pi^0$, $\bar {p}K^{0}_{S}\pi^+\pi^-$, $\bar\Lambda\pi^-$, $\bar\Lambda\pi^-\pi^-\pi^+$, $\bar\Lambda\pi^-\pi^0$, $\bar{\Sigma}^0\pi^-$ and $\bar{\Sigma}^-\pi^+\pi^-$. These reconstructed decays are referred to as single-tag (ST) $\bar\Lambda^{-}_{c}$ baryons, where the intermediate particles
$K^{0}_{S}$, $\bar{\Lambda}$, $\bar{\Sigma}^0$, $\bar{\Sigma}^-$, and $\pi^0$ are reconstructed via $K^{0}_{S}\to \pi^+\pi^-$, $\bar{\Lambda}\to\bar{p}\pi^+$, $\bar{\Sigma}^0\to\gamma\bar{\Lambda}$, $\bar{\Sigma}^-\to\bar{p}\pi^0$ and $\pi^0\to\gamma\gamma$, respectively.
In the other side of the events recoiling against the ST $\bar\Lambda^{-}_{c}$ baryons, the candidate $\Lcsg$ decays are selected to
form double-tag events (DT).

\section{ST event selection}
The same selection criteria  are used in this analysis as in Ref.~\cite{p3}. Charged tracks are required to have a polar angle ($\theta$) within $|\!\cos\theta| < 0.93$, where $\theta$ is defined with respect
to the beam direction. Except for those from $K^0_S$ and $\bar{\Lambda}$ decays, all tracks are required to originate from an interaction region defined by $|V_{xy}| < 1$~cm and $|V_z| < 10$~cm, where $|V_{xy}|$ and $|V_z|$ refer to the distances of closest approach of the reconstructed track to the interaction point (IP) in the $xy$ plane and the $z$ direction (along the beam), respectively.

Particle identification (PID) is implemented
by combining the measurements of ${\rm d}E/{\rm d}x$ in the MDC and the flight time in the TOF into a probability that a given track is a pion, kaon or proton.  The track is assigned to one of these three particle types, according to the probability.

Candidates for $K^{0}_{S}$ and $\bar{\Lambda}$ mesons are reconstructed from their decays to $\pi^+\pi^-$ and $\bar{p}\pi^+$, respectively, where the charged tracks must have
distances of closest approach to the IP that are within $\pm$20~cm along the beam direction.
To improve the signal purity, PID is implemented to select the antiproton candidate,
while the charged pion is not subject to any PID requirement.
A secondary vertex fit is performed for each $K^{0}_{S}$ or $\bar{\Lambda}$ candidate,
and the momenta updated by the fit are used in the subsequent analysis. The $K^{0}_{S}$ or
$\bar{\Lambda}$ candidate is accepted if the $\chi^2$ of the secondary vertex fit is less than 100.
Furthermore, the decay vertex is required to be separated from the IP by a distance of at least twice the fitted vertex resolution, and the invariant mass to be
within (0.487, 0.511)~GeV/$c^2$ for $\pi^+\pi^-$  or (1.111, 1.121)~GeV/$c^2$ for the $\bar{p}\pi^+$ pair. The two invariant mass for $\pi^+\pi^-$ and $\bar{p}\pi^-$ resolutions are found, using MC simulations, to be 2.9 MeV$/c^{2}$ and 1.2 MeV$/c^{2}$, respectively. The $\bar{\Sigma}^0$ and $\bar{\Sigma}^-$ candidates are reconstructed from the $\gamma\bar{\Lambda}$
and $\bar{p}\pi^0$ final states with invariant masses being within
(1.179, 1.203)~GeV/$c^2$ and (1.176, 1.200)~GeV/$c^2$, respectively. The two invariant mass resolutions are found, using simulation, to be 3.6 MeV$/c^{2}$ and 4.3 MeV$/c^{2}$, respectively.

Photon candidates are identified using showers in the electromagnetic calorimeter (EMC). The deposited energy of each shower must be
more than 25~MeV in the barrel region ($|\!\cos\theta| \le 0.80$) or more than 50~MeV in the end-cap region
($0.86 \le |\!\cos\theta| \le 0.92$). To suppress electronic noise and showers unrelated to the event,
the difference between the EMC time and the event start time is required to be within
[0, 700]~ns.

The $\pi^0$ candidates are reconstructed with pairs of photon candidates within the invariant mass range
(0.115, 0.150)~GeV/$c^2$. To improve the resolution, a kinematic fit is performed by constraining the invariant
mass of the photon pair to correspond to the $\pi^0$ mass and requiring the corresponding $\chi^2$ of the fit to be less than 200.
The momenta updated by the kinematic fit are used in the subsequent analysis.

To distinguish the ST $\bar\Lambda^{-}_{c}$ baryons from combinatorial backgrounds, we study the distributions of the energy difference $\Delta E$ and the beam-constrained mass $M_{\rm BC}$ of the selected ST candidates,
 defined as
\begin{equation}
\Delta E\equiv E_{\bar\Lambda^{-}_{c}}-E_{\mathrm{beam}},
\end{equation}
\begin{equation}
M_{\rm BC}\equiv\sqrt{E_{\mathrm{beam}}^{2}/c^{4}-|\vec{p}_{\bar\Lambda^{-}_{c}}|^{2}/c^{2}},
\end{equation}
\noindent where $E_{\rm beam}$ is the beam energy, and $\vec{\mkern1mu p}_{\bar\Lambda^{-}_{c}}$ and $E_{\bar\Lambda^{-}_{c}}$ are the total momentum and energy of the ST candidate, respectively, calculated in the $e^+e^-$ rest frame. The signals are expected to concentrate around zero in $\Delta E$ distribution and around the nominal $\bar\Lambda^{-}_{c}$ mass in the $M_{\rm BC}$ distribution. If there are multiple candidates for each tag mode, the one with minimum $|\Delta E|$ is retained. Combinatorial backgrounds in the $M_{\rm BC}$ distributions are suppressed with the $\Delta E$ requirements shown in Table~\ref{tab:yield-st-460}.

\begin{figure}[!htp]
    \begin{center}
     \includegraphics[width=0.42\textwidth]{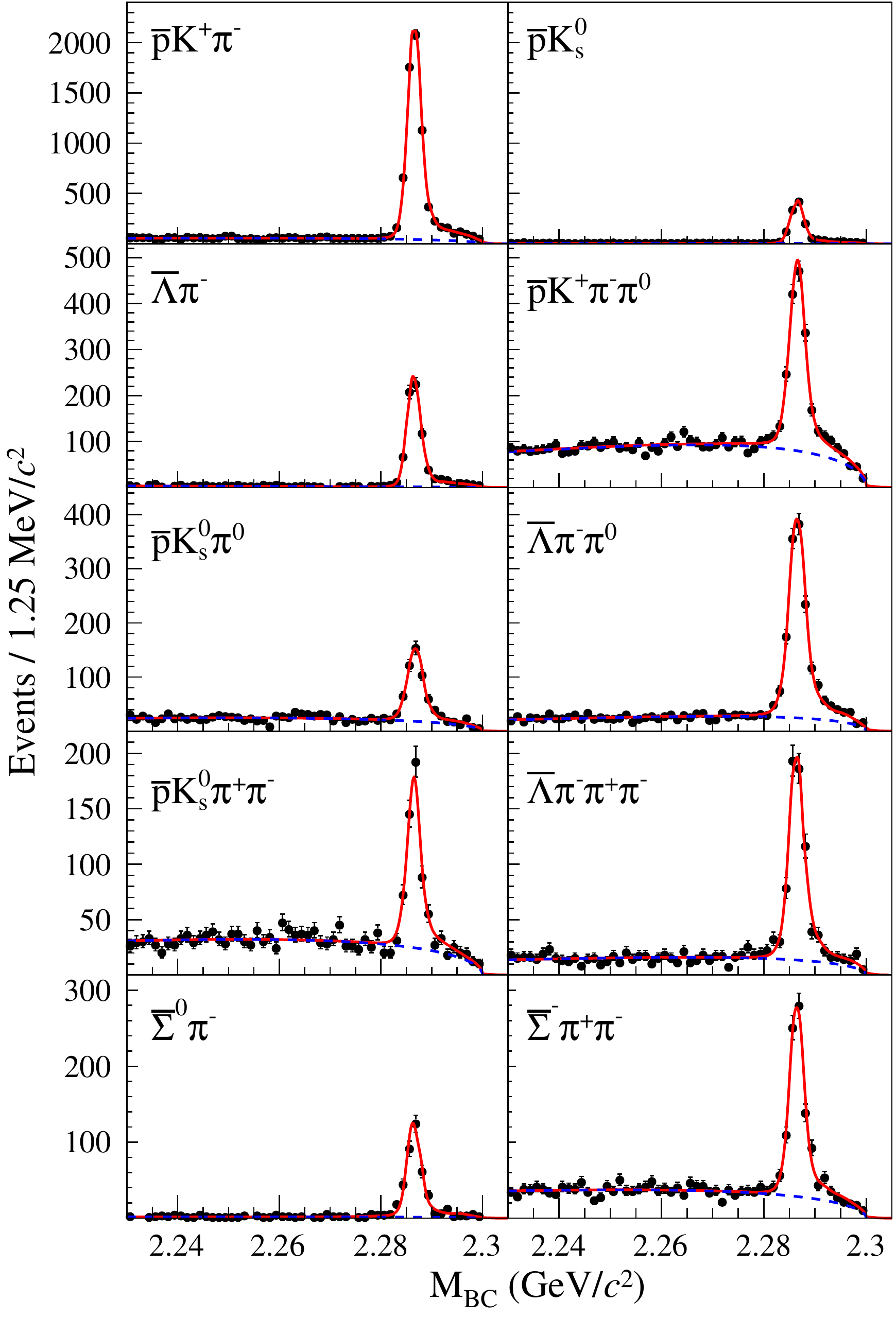}
    \end{center}
    \caption{
      Distributions of $\mbc$ for the different ST channels of the $\LamCB$ at $\sqrt{s}=4.600$ GeV. The signal shape of the $\LamCB$ is described
      by the simulated shape convolved with a Gaussian resolution function and the background is modeled
      with an ARGUS function. The points with error bars represent
data. The (red) solid curves indicate the fit results and the
(blue) dashed curves describe the background shapes.} \label{fig:single-tag-460}
\end{figure}

    For each tag mode, the ST yield is determined by fitting the $M_{\rm BC}$ distribution of the candidates accepted by all the  requirements described. In the fit, the $\LamCB$ signal is modeled with a shape obtained from the MC-simulated signal convolved with Gaussian function, different for different energies and tagging modes, and the combinatorial background is described by an ARGUS function~\cite{argus}. The fits to the $M_{\rm BC}$ distributions for the various tag modes at $\sqrt{s}=4.600 \gev$ are shown in Fig.~\ref{fig:single-tag-460}. Candidates in the $M_{\rm BC}$ signal region, $(2.275, 2.310)$~GeV$/c^2$, are kept for further analysis. The ST yields in data and the ST efficiencies for individual tags are
 shown in Table~\ref{tab:yield-st-460}. The same procedure is performed for the other six data
 samples at different energy points which are summarized in the Supplementary materials of Ref.~\cite{p3}. Summing over the ST yields for all tags and energy points gives the total ST yield to be $N^{\rm tot}_{\rm ST}=105244 \pm 384$,  where the uncertainty is statistical.

\begin{table}[!htbp]
  \begin{center}
  \caption{The $\Delta E$ requirement, ST yield, and ST detection efficiency (include the branching fractions of subleading decays) of $\bar\Lambda^{-}_{c} \to \bar\Sigma^{-} \gamma$
           for each tag mode for the data sample at $\sqrt{s}=4.600$~GeV. The uncertainty in the ST yield is statistical only. }
    \begin{tabular}{ l c r @{ $\pm$ } p{1cm} c}
      \hline
      \hline
         Tag mode   & $\Delta E$(MeV) & \multicolumn{2}{c}{$N_{i}^{\mathrm{ST}}$}  & $\epsilon_{i}^{\mathrm{ST}}$(\%)    \\
      \hline

            $\Bpkpi$                       & $(-34,~20)$    &  $6705$     & 90  &  51.0   \\
            $\Bpks$                        & $(-20,~20)$    &  $1268$     & 37  &  56.2   \\
            $\bar{\Lambda}\pi^-$           & $(-20,~20)$    &  $741$      & 28  &  47.7   \\
            $\Bpkpi\pi^0$                  & $(-30,~20)$    &  $1539$     & 57  &  15.4   \\
            $\Bpks\pi^0$                   & $(-30,~20)$    &  $485$      & 29  &  18.4   \\
            $\bar{\Lambda}\pi^-\pi^0$      & $(-30,~20)$    &  $1382$     & 49  &  16.6   \\
            $\Bpks\pi^+\pi^-$              & $(-20,~20)$    &  $512$      & 29  &  19.9   \\
            $\bar{\Lambda}\pi^-\pi^+\pi^-$ & $(-20,~20)$    &  $646$      & 31  &  13.7   \\
            $\bar{\Sigma}^0\pi^-$          & $(-20,~20)$    &  $404$      & 22  &  22.5   \\
            $\bar{\Sigma}^-\pi^+\pi^-$     & $(-30,~20)$    &  $872$      & 38  &  18.1   \\
      \hline\hline
    \end{tabular}
    \label{tab:yield-st-460}
  \end{center}
\end{table}

\section{DT event selection}

After the selection of the tag side, the $\Lambda_{c}^+\to \Sigma^+\gamma$ ($\Sigma^{+} \rightarrow p\pi^0$) is selected in the recoil side of the tagged $\LamCB$ as follows.
It is required to have only one good charged track identified as a proton, apart from those charged tracks used in the ST selection.  The proton candidate is required to originate from within $20$ cm along the beam axis with respect to the IP. The $\pi^0$ candidates are selected with photon pairs, and the energies of the photons are required to be less than 0.45 GeV,  which is a requirement set from the study of the  $\pi^{0} \to \gamma \gamma$ distribution in signal MC at generator level.
  The energy of the radiative isolated high energy $\gamma$ is required to be greater than 0.65 GeV, and
  the number of good photons must be exactly one for the signal process.
  For multiple $\pi^0$ candidates, only the combination of the proton, $\pi^0$ and
  radiative $\gamma$ with minimum of $|\Delta E_{\rm sig}|$ is retained, where
  $\Delta E_{\rm sig}$ = $E_{p\pi^0 \gamma}-E_{\mathrm{beam}}$.
  The $\Delta E_{\rm sig}$ is further required to be within $(-0.038, 0.026)$ GeV.
    The $\Sigma^{+}$ candidate is reconstructed via $\Sigma^{+} \rightarrow p\pi^0$ with a mass ($M_{p\pi^{0}}$) lying within (1.176, 1.200)~GeV/c$^2$.

To determine the detection efficiency, 500,000 events of $\Lcsg$ are simulated for each energy point. The ten tag modes are simulated according to their relative BFs.  The  DT efficiencies measured from this MC sample are summarized in Table~\ref{dteff} after all selection requirements.

\begin{table}[htbp]
  \begin{center}
  \caption{The DT detection efficiency (include the branching fractions of subleading decays) in percent for each tag mode and each energy point.}\label{dteff}
  \renewcommand\arraystretch{1.8}
    \begin{tabular}{ l c c c c c c c}
      \hline
      \hline
            $\sqrt{s}$ (GeV) & 4.600 & 4.612 & 4.628 & 4.641 & 4.661 & 4.682 & 4.699 \\
      \hline
         $\pkpi$                   &  12.9   &  12.7    &  12.2    &  11.9  & 11.6 & 11.2  & 10.9   \\
         $\pks$                    &  13.8   &  13.2    &  12.7   &   12.0  & 11.6 & 11.2 & 11.0  \\
         $\Lambda\pi^+$            &  10.4     &10.0    &  9.6    &   9.1   & 8.8 & 8.7 & 8.0 \\
         $\pkpi\pi^0$              &  4.4    & 4.3    &  4.1     &   3.9   & 3.8 & 3.8 & 3.6 \\
         $\pks\pi^0$               &  4.7     & 4.6    & 4.4      &   4.2  & 4.1 & 4.0 & 3.9 \\
         $\Lambda\pi^+\pi^0$       &  4.1   &  3.7    &  3.7     &   3.7   & 3.4 & 3.5 & 3.3 \\
         $\pks\pi^+\pi^-$          &  5.2    &  4.8    & 4.6     &   4.7   & 4.4 & 4.3 & 4.1 \\
         $\Lambda\pi^+\pi^+\pi^-$  &  3.6    & 3.4    &  3.3     &   3.1   & 3.2 & 3.1 & 3.1 \\
         $\Sigma^0\pi^+$           &  6.0    & 5.5    &  5.1     &   5.4   & 4.8 & 4.8 & 4.8 \\
         $\Sigma^+\pi^+\pi^-$      &  4.7     & 4.3    &  4.3    &   4.2   & 4.3 & 3.9 & 3.9 \\
      \hline\hline
    \end{tabular}
  \end{center}
\end{table}

\section{Background analysis}
Potential sources of background are classified into two categories: those directly originating from continuum hadron production
in $\ee$ annihilation (denoted as $q\bar{q}$ background thereafter) and those from $\ee$ $\to$ $\LCpair$ (denoted as $\LCpair$ background thereafter).

The $M^{\rm sig}_{\rm BC}$ distribution of the accepted candidates from the $q\bar q$ component in the inclusive MC sample is shown in Fig.~\ref{fig:limit}(a), where no peaking contribution in the signal region is observed.
The yield of the $q\bar q$ component in the signal region $M_{\rm BC}^{\rm sig}\in (2.275, 2.310)$ GeV/$c^2$   is estimated from the data  to be $5.0 \pm 0.2$ events, which is determined by measuring the number of data events in the sideband region $M_{\rm BC}^{\rm sideband}\in (2.15, 2.27)$ GeV/$c^2$. The $M_{\rm BC}^{\rm sig}$ distributions in the inclusive MC sample show that the background components are consistent between sideband and signal regions.
This sideband contribution has been extrapolated to the signal region with a scale  factor 1.03, which is the ratio between sideband region and signal region in the inclusive MC sample.

The $\LCpair$ background is dominated by events containing the decay  $\Lambda^{+}_{c} \to \Sigma^{+} \pi^0$,
the magnitude of which is estimated using MC simulation.
The expected background yield from this source is estimated in MC simulation and found to be $6.2 \pm 0.5$ events after normalizing to the integrated luminosity of the data sample and  taking
${\mathcal B}(\Lambda^{+}_{c} \to \Sigma^{+} \pi^{0}) = (1.24 \pm 0.10)\times10^{-2}$
from the PDG~\cite{g3}.
The resulting $M_{BC}$ distribution of the accepted candidates in data, signal MC and various simulated background contributions is presented in Fig.~\ref{fig:limit}(a), where the signal MC is shown with a BF = $1.0\times10^{-4}$, which corresponds to an event yield of 1.2 events. There are 10 events  in data in the signal region.

\section{Upper limit calculation}
\begin{figure}[!htp]
\begin{center}
\subfigure[]{\includegraphics[width=0.43\textwidth,height=0.32\textwidth]{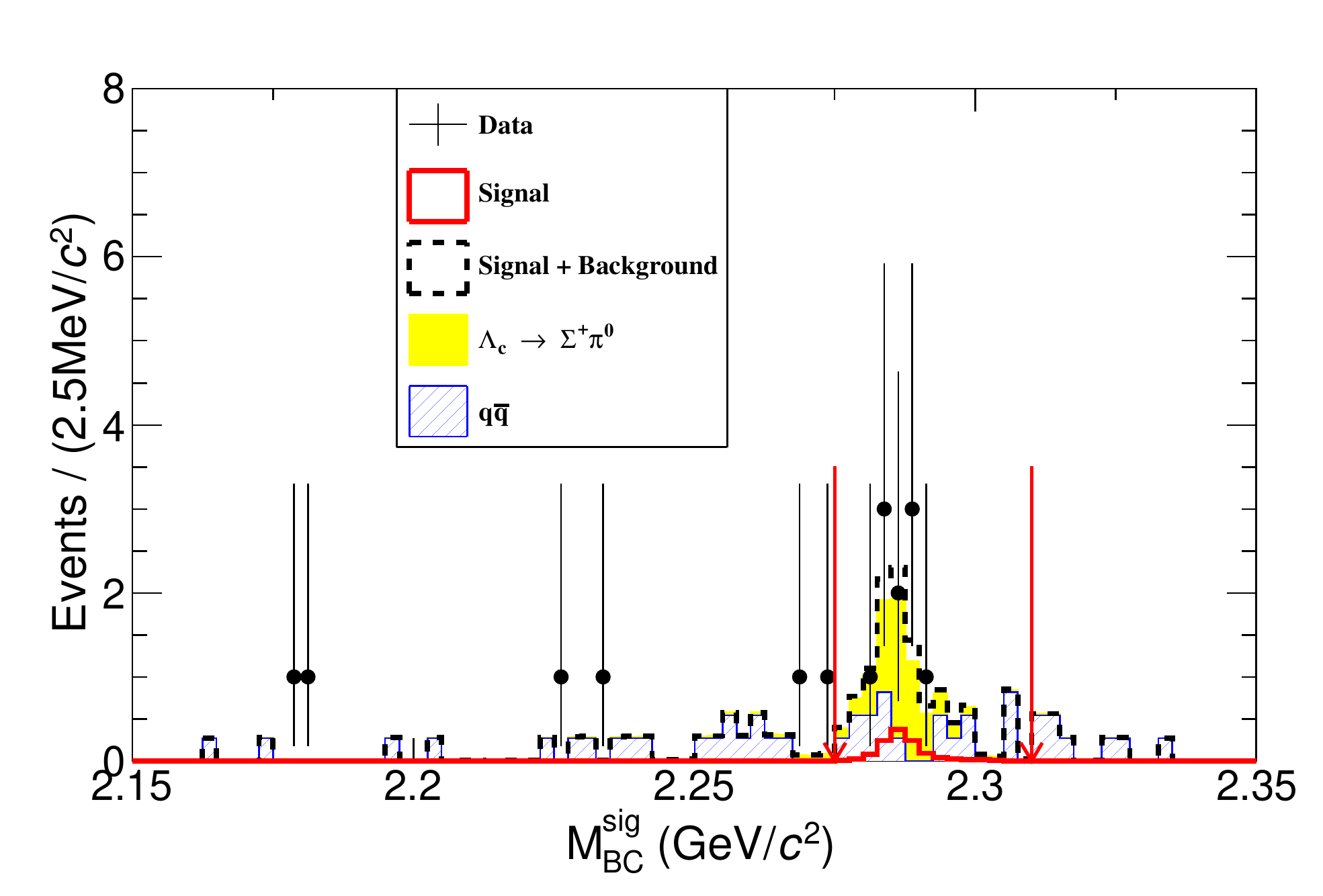}}
\subfigure[]{\includegraphics[width=0.45\textwidth,height=0.3\textwidth]{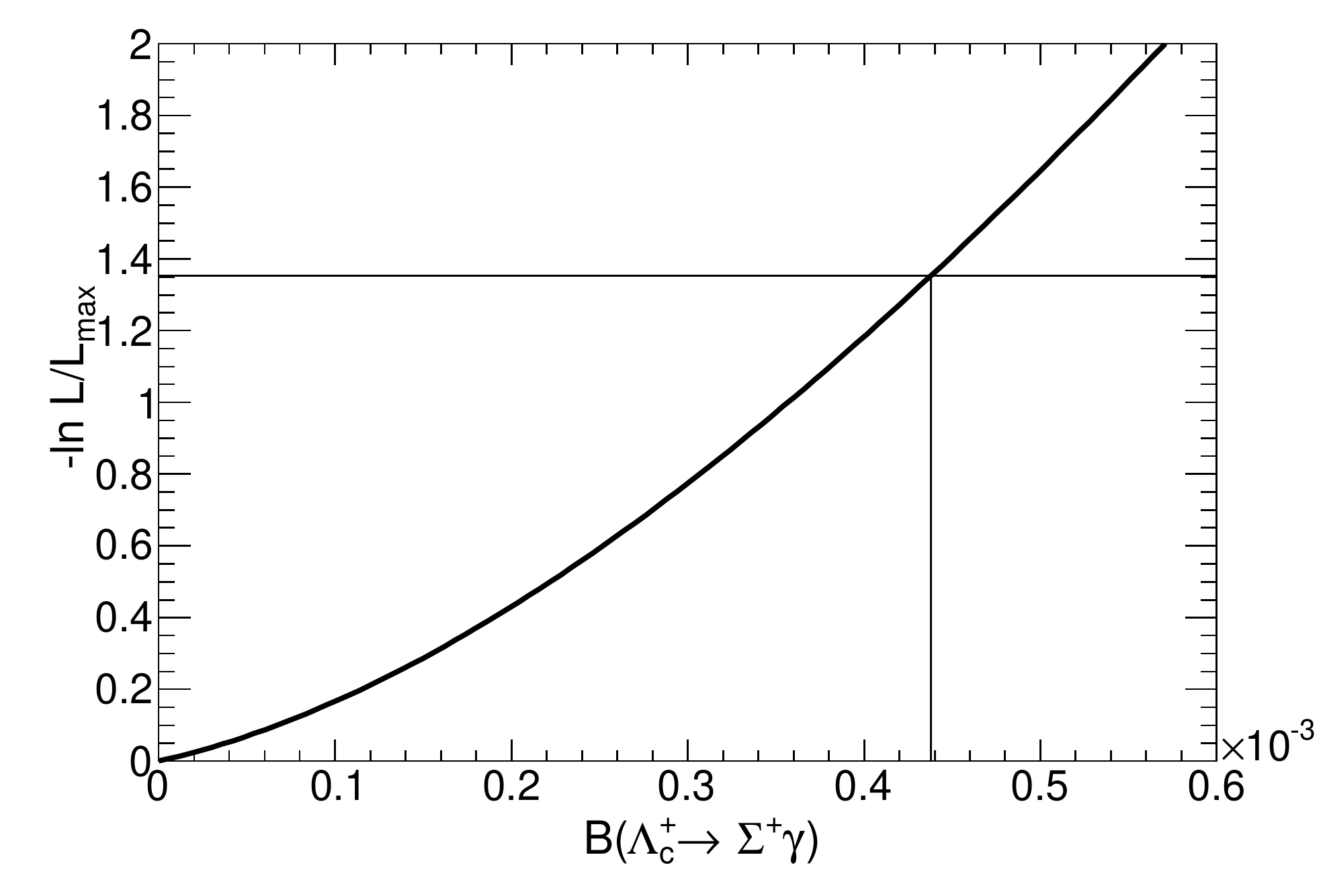}}
\caption{(a) The comparison of the $\mbc^{\rm sig}$ distributions of the candidate events for $\Lcsg$ between data and MC simulation.
The two red arrows indicate the signal region.
  (b) The profile log-likelihood ratio curve versus ${\mathcal B}(\Lcsg)$.
   The intersection of the curve and horizontal line indicates the upper limit of the BF at the 90\% CL, where the black solid curve
   is the scan result with systematic uncertainties.
\label{fig:limit}
}
\end{center}
\end{figure}

Since no significant signal is observed, a test statistic based on a profile log-likelihood ratio~\cite{lim} is used to determine
the upper limit on the BF of $\Lcsg$. The likelihood function depending on the parameter of interest
${\mathcal B}(\Lcsg)$ and the nuisance parameters $\theta_{1} = (\epsilon_{\rm eff},N_{\rm bkg})$
is defined as
\begin{equation}
{\mathcal L}(\mathcal B(\Lcsg),\theta_{1}) = {\rm Pois}(N_{\rm obs}|N_{\rm exp})\cdot {\rm Gaus}(\theta_{1}),
\end{equation}
where Pois is a Poisson function, Gaus is a Gaussian distribution, $N_{\rm obs}$ is the number of events observed in the signal region from data. $N_{\rm exp}$ is the expected number of events, which is defined as the sum of the number of background and signal events estimated in the signal region,
\begin{equation}
N_{\rm exp} = 2 N_{\Lambda^{+}_{c}\bar{\Lambda}^{-}_{c}}\cdot\mathcal{B}_{\rm ST}\cdot\mathcal B(\Lcsg)\cdot\epsilon_{\rm eff}+N_{\rm bkg},
\end{equation}
where $N_{\Lambda^{+}_{c}\bar{\Lambda}^{-}_{c}}$ is the number of $N_{\Lambda^{+}_{c}\bar{\Lambda}^{-}_{c}}$ pairs, $\mathcal{B}_{\rm ST}$ stands for the BF of a given tag mode, $\epsilon_{\rm eff}$ refers to signal efficiency and $N_{\rm bkg}$ is the number of background events.
The systematic uncertainties are parameterized as Gaussian-function constraints with nuisance parameters $\theta_{1}$. The uncertainties associated with the efficiency and background  are considered separately.
 The upper limit on the BF at a confidence level (CL) of 90$\%$ is derived by scanning the parameter of interest space, with result shown in Fig.~\ref{fig:limit}(b), and is found to be
 ${\mathcal B}(\Lcsg) < 4.4\times10^{-4}$.

\section{Systematic uncertainties}
With the DT method, most of the systematic uncertainties arising from the ST side cancel out. However,  systematic uncertainties originate from other sources. The uncertainty in the total ST yield is assigned as 0.5\%~\cite{g1}, which comes  mainly from the statistical uncertainty, with  an additional systematic component associated with the fit to the $M_{BC}$ distribution of the ST candidates, assessed by varying the signal
shape, the background shape, and the fit range in the fit to the $M_{BC}$ distributions. The uncertainties due to the proton tracking and PID efficiencies are studied with a control sample of reconstructed $J/\psi \to p\bar{p} \pi^+\pi^-$ process. These are assigned to be 2.0\% and 1.0\%, respectively. The uncertainties associated with the difference in  $\pi^{0}$ reconstruction efficiencies between data and MC simulation are estimated using DT events with $\bar D^0\to K^+\pi^-$, $K^+\pi^-\pi^-\pi^+$ versus $D^{0} \to K^{-}\pi^{+}\pi^{0}$ . The systematic uncertainty in the reconstruction efficiency per radiative photon is based on the studies with the control sample of $J/\psi \to \rho^{0}\pi^{0}$ with $\rho^{0} \to \pi^{+}\pi^{-}$ and $\pi^{0} \to \gamma \gamma$~\cite{gamerr} and is assigned to 1.0\%. The difference of $\pi^0$ efficiencies between data and MC simulation, 1.0\%, is assigned as the associated systematic uncertainty~\cite{pi0err}. The uncertainty due to the requirement on only one proton candidate is estimated from a sample of  $J/\psi \to \Sigma^0 \bar{\Sigma^0} $ ($\Sigma^0 \to p^+ \pi^- $) decays. The efficiency difference between this requirement and the nominal one is 2.0\%, which is assigned as a systematic uncertainty.    The potential bias due to the $M_{p\pi^{0}}$ requirement is estimated with the $J/\psi \to \Sigma^{+} \Sigma^- $ sample, with the 0.4\% difference of efficiencies between data and simulation being assigned as the corresponding uncertainty. The systematic uncertainty due to the $\Delta E_{\rm sig}$ requirement is estimated in a similar manner with the $\Lambda_{c} \to p \pi^0 \pi^0$ sample and found to be 3.1\%. The potential bias associated with the $q\bar{q} \to$~hadrons background contribution
is assigned from the
the statistical uncertainty of the data yield in the sideband range of the $M_{\rm BC}^{\rm sig}$ distribution (2.15, 2.27)~GeV/$c^2$, which is used to estimate this background, and leads to a  4.4\% uncertainty in the branching fraction. The total systematic uncertainty from the $q\bar{q} \to$~hadrons background contamination is assigned to be 2.1\% as the quadratic sum of the above two sources. The systematic uncertainty associated with the $\Lambda_{c} \to \Sigma \pi^0$ background contribution is 4.4\% based on the knowledge of the BF from PDG~\cite{g3}. To study the effects of the uncertainty in the MC model we generate the $\Lcsg$ events with a proton polar-angle distribution parameterized by 1 + $\alpha \cos^2\theta$ (with $\alpha = \pm 1.0$), and find a difference of 2.1\% in efficiency with respect to the baseline phase-space model.  The limited size of the signal MC sample leads to a  0.6\% uncertainty in the knowledge of the efficiency. Table~\ref{tab:sys-sum} summarizes the sources of the systematic uncertainties in the measurement of branching ratio of
$\Lcsg$ decay. The total systematic uncertainty is obtained to be 7.3\% as the quadratic sum of all sources.

\begin{table}[htbp]
  \begin{center}
          \caption{Summary of relative systematic uncertainties (in \%). }\label{tab:sys-sum}
  \renewcommand\arraystretch{1.8}
    \begin{tabular}{ l c c c}
      \hline
      \hline
	    Sources &  $\mathcal{B}(\Lcsg)$ (\%)\\
      \hline
	    Single-tag yield                     & 0.5  \\
	    $p$ tracking                         & 2.0  \\
	    $p$ PID                              & 1.0  \\
        $\gamma$ detection                   & 1.0  \\
	    $\pi^{0}$ reconstruction             & 1.0 \\
        $N_{\rm proton}$ requirement    & 2.0 \\
        Signal model                           & 2.1 \\
        $\Sigma^{+}$ requirement              & 0.4  \\
	    $\dE$ requirement                     & 3.1  \\
        $q\bar{q}$ background                  & 2.1   \\
        $\Lambda_{c} \to \Sigma \pi^0$ background  & 4.4   \\
        Assumed BF (PDG)                       & 0.6 \\
	    MC statistics                        & 0.6  \\
       \hline
        Sum                                  & 7.3 \\
      \hline\hline
    \end{tabular}
  \end{center}
\end{table}

\section{Summary}
Using an $e^+e^-$ collision data sample corresponding to an integrated luminosity of 4.5 $\rm fb^{-1}$ collected at $\sqrt s=4.60 - 4.70$~GeV with the BESIII detector,
we have searched for the Cabibbo-allowed weak radiative decay $\Lcsg$ in a model-independent approach for the first time.
 No signal is found, and  an upper limit on the BF of $\Lcsg$ decay is set to be $4.4\times10^{-4}$ at the 90\% CL. This result is consistent with the theoretical predictions of $5\times10^{-5}$~\cite{08}, $2.8\times10^{-4}$~\cite{qcd} and $1.03\times10^{-4}$~\cite{light}, from the Bag model and appropriate QCD corrections, respectively, where the short-distance $cd \to u s \gamma$ mechanism is expected to be dominant. A more stringent constraint, or discovery, is expected with the larger data set that BESIII expects to accumulate in the near future~\cite{white}.

\section{Acknowledgement}
The BESIII collaboration thanks the staff of BEPCII and the IHEP computing center for their strong support. This work is supported in part by National Key R\&D Program of China under Contracts Nos. 2020YFA0406300, 2020YFA0406400; National Natural Science Foundation of China (NSFC) under Contracts Nos. 11635010, 11735014, 11835012, 11935015, 11935016, 11935018, 11961141012, 12022510, 12025502, 12035009, 12035013, 12192260, 12192261, 12192262, 12192263, 12192264, 12192265,  12005311; the Fundamental Research Funds for the Central Universities, Sun Yat-sen University, University of Science and Technology of China; 100 Talents Program of Sun Yat-sen University; the Chinese Academy of Sciences (CAS) Large-Scale Scientific Facility Program; Joint Large-Scale Scientific Facility Funds of the NSFC and CAS under Contract No. U1832207; the CAS Center for Excellence in Particle Physics (CCEPP); 100 Talents Program of CAS; The Institute of Nuclear and Particle Physics (INPAC) and Shanghai Key Laboratory for Particle Physics and Cosmology; ERC under Contract No. 758462; European Union's Horizon 2020 research and innovation programme under Marie Sklodowska-Curie grant agreement under Contract No. 894790; German Research Foundation DFG under Contracts Nos. 443159800, Collaborative Research Center CRC 1044, GRK 2149; Istituto Nazionale diFisica Nucleare, Italy; Ministry of Development of Turkey under Contract No. DPT2006K-120470; National Science and Technology fund; National Science Research and Innovation Fund (NSRF) via the Program Management Unit for Human Resources \& Institutional Development, Research and Innovation under Contract No. B16F640076; Olle Engkvist Foundation under Contract No. 200-0605; STFC (United Kingdom); Suranaree University of Technology (SUT), Thailand Science Research and Innovation (TSRI), and National Science Research and Innovation Fund (NSRF) under Contract No. 160355; The Royal Society, UK under Contracts Nos. DH140054, DH160214; The Swedish Research Council; U. S. Department of Energy under Contract No. DE-FG02-05ER41374.

\end{document}